\documentclass[english]{article}
\usepackage[utf8]{inputenc}
\usepackage{natbib}
\usepackage[OT1]{fontenc}

\usepackage{geometry}
\usepackage{babel}
\usepackage{float}
\usepackage{mathtools}
\usepackage{bm}
\usepackage{bbm}
\usepackage{amsmath}
\usepackage{amssymb}
\usepackage{dsfont}
\usepackage{graphicx}
\usepackage[unicode=true,pdfusetitle,
 bookmarks=true,bookmarksnumbered=false,bookmarksopen=false,
 breaklinks=false,pdfborder={0 0 1},backref=false,colorlinks=false]
 {hyperref}

\makeatletter

\floatstyle{ruled}
\newfloat{algorithm}{tbp}{loa}
\providecommand{\algorithmname}{Algorithm}
\floatname{algorithm}{\protect\algorithmname}

\usepackage{babel}

\usepackage{algorithmic}

\usepackage{amsthm}
\usepackage{enumitem}

\theoremstyle{plain}

\newtheorem{lemma}{\textbf{Lemma}}
\newtheorem{theorem}{\textbf{Theorem}}

\newtheorem{assumption}{\textbf{Assumption}}
\newtheorem{definition}{\textbf{Definition}}

\theoremstyle{definition}
\newtheorem{remark}{\textbf{Remark}}

\newtheorem{example}{\textbf{Example}}

\usepackage{color}
\definecolor{cm}{RGB}{0,0,200}

\definecolor{yy}{RGB}{160, 32, 240}

\usepackage{dsfont}

\usepackage{authblk}

\author{Kaining Shi} \author{Cong Ma} \affil{Department of Statistics, University of Chicago} 

\@ifundefined{showcaptionsetup}{}{%
 \PassOptionsToPackage{caption=false}{subfig}}
\usepackage{subfig}
\makeatother

\begin{document}
\title{Auditing Differential Privacy in the Black-Box Setting}
\maketitle
\begin{abstract}
This paper introduces a novel theoretical framework for auditing differential privacy (DP) in a black-box setting. Leveraging the concept of $f$-differential privacy, we explicitly define type I and type II errors and propose an auditing mechanism based on conformal inference. Our approach robustly controls the type I error rate under minimal assumptions. Furthermore, we establish a fundamental impossibility result, demonstrating the inherent difficulty of simultaneously controlling both type I and type II errors without additional assumptions. Nevertheless, under a monotone likelihood ratio (MLR) assumption, our auditing mechanism effectively controls both errors. 
We also extend our method to construct valid confidence bands for the trade-off function in the finite-sample regime.
\end{abstract}

\section{Introduction}\label{sec: intro} 
In the era of big data, privacy concerns have become a critical issue in numerous real-world applications, including healthcare, finance, and social networks. As organizations collect and analyze vast amounts of sensitive user data, it is crucial to ensure that individual privacy is preserved while still enabling meaningful data-driven insights. Differential Privacy (DP) has emerged as a rigorous mathematical framework and widely adopted by organizations such as the U.S. Census Bureau (\cite{USCensus}), as well as major companies like Apple (\cite{Apple}), Google (\cite{Google}), and Microsoft (\cite{Microsoft}). By introducing carefully calibrated noise to data queries or models, DP ensures that the presence or absence of any single individual in a dataset does not significantly affect the overall output, thereby preventing privacy breaches.  

Despite its theoretical guarantees, the practical deployment of DP mechanisms necessitates thorough auditing to verify whether a system adheres to the intended privacy standards. Differential privacy auditing serves as a crucial tool to evaluate and quantify the actual privacy loss of a system, ensuring compliance with established privacy parameters. Effective auditing mechanisms help bridge the gap between theoretical DP models and their real-world implementations, where factors such as incorrect noise calibration, unexpected data processing steps, or adversarial attacks may undermine the intended privacy protections.  

While DP auditing can be performed in both white-box and black-box settings, the black-box auditing framework holds particular significance in practice. In many real-world scenarios, organizations deploy DP mechanisms within proprietary systems, limiting access to internal implementation details. Black-box auditing enables privacy verification without requiring full knowledge of the system’s inner workings, making it a more realistic and applicable approach in environments where transparency is constrained. Furthermore, black-box auditing aligns with adversarial settings, where privacy attacks are conducted without internal access, making it a valuable tool for assessing the robustness of DP mechanisms against potential threats.  

Given the increasing adoption of DP in privacy-preserving applications, the development of effective black-box auditing techniques is essential. This paper explores the possibilities and impossibilities of differential privacy auditing in the black-box setting and provides a novel theoretical framework for black-box auditing of differential privacy from a non-asymptotic perspective.

\paragraph{Main contributions.}
This work makes the following contributions. 

\begin{itemize}
\item We introduce a new conceptual framework for auditing differential privacy in black-box scenarios, explicitly capturing the adversarial nature inherent to privacy auditing.

\item We propose a practical, computationally efficient auditing algorithm leveraging conformal inference principles, capable of robustly controlling the type I error in the finite-sample case without requiring restrictive assumptions.

\item We identify and rigorously prove an impossibility result that underscores the fundamental limitations in achieving simultaneous control over type I and type II errors.

\item Under a general monotone likelihood ratio (MLR) assumption, we demonstrate that our auditing mechanism can effectively achieve simultaneous control over both error types.

\item We extend our methodology to construct a valid confidence band for the trade-off function in the black-box setting that has finite-sample validity. 
\end{itemize}

\subsection{Related work}
Since its introduction in \cite{DP-Definition}, differential privacy has been widely adopted by major companies and government agencies such as Apple(\cite{Apple}), Google (\cite{Google}), Microsoft (\cite{Microsoft}), and the U.S. Census Bureau (\cite{USCensus}), exerting a significant influence. For a long time, research primarily focused on exploring the implementation of differential privacy in white-box algorithms, such as bandit algorithms (\cite{guha2013nearly}, \cite{shariff2018differentially}, \cite{DPBandit}, \cite{hanna2022differentially},\cite{charisopoulos2023robust}), deep learning (\cite{deeplearning}, \cite{deeplearning2}, \cite{deeplearning3}), or neural network (\cite{neuralnetwork1}, \cite{neuralnetwork2}).

However, considering that users or regulators may be more concerned about whether the algorithm can guarantee the privacy of its users, there has been growing interest in the issue of differential privacy auditing in recent years. Since users or regulatory authorities often lack a white-box understanding of the underlying algorithms—especially with the growing complexity of modern algorithms, such as the use of large language models to achieve differential privacy (\cite{DP-LLM})—it is becoming increasingly difficult to have a clear white-box perception of these algorithms. It may even be challenging to make any assumptions about the outputs of such complex algorithms. In this context, there is a growing interest in privacy auditing under black-box settings.

The existing literature provides various empirical methods (\cite{Auditing1}, \cite{Auditing2}). \cite{Auditing3} proposes an auditing framework with certain statistical guarantees; however, it still requires the empirical selection of the event $E$ in the definition of classical $(\varepsilon,\delta)$-DP (\cite{DP-Definition}). \cite{Auditing-Impossibility} leverages the discontinuity in the definition of differential privacy, arguing that under two sufficiently close output distributions of an algorithm, it is possible to reach two drastically different conclusions—satisfying or violating differential privacy—demonstrating the impossibility of complete differential privacy auditing with finite samples. Most of the existing studies adopt the framework based on the definition of classical $(\varepsilon,\delta)$-DP (\cite{DP-Definition}). In contrast, \cite{GDP} offers a new perspective by introducing the differential privacy function, which facilitates a more comprehensive and general discussion on differential privacy auditing.

Based on Definition \ref{definition-f-DP}, some existing studies have proposed empirical (\cite{empirical-auditing-f-dp-1}, \cite{empirical-auditing-f-dp-2}) or asymptotically guaranteed auditing methods (\cite{DPAuditing}). Contemporaneous work \cite{DPAuditing} also investigates the auditing of f-DP under the black-box setting. Their idea is similar to \cite{Auditing-density-estimation}, obtaining asymptotic guarantees by estimating the densities of two distributions. These asymptotic results rely on fixed distribution pairs and assume that the sample size approaches infinity. However, a fundamental difference is that we argue that for privacy auditing under the black-box setting, a more appropriate framework might be “adversarial.” Concretely, when the mechanism and sample size are fixed, we need to examine our mechanism under the “worst-case” scenario. In this context, obtaining meaningful asymptotic results is challenging (Theorem \ref{theorem-Impossibility}).

\section{Background and Problem Formulation}
In this section, we briefly review the $f$-differential privacy ($f$-DP) framework introduced in \cite{GDP}, followed by a formal statement of the auditing problem we consider.

\subsection{\texorpdfstring{$f$-Differential Privacy}{f-Differential Privacy}}

Differential privacy is closely related to statistical hypothesis testing. 
Consider the hypothesis testing problem 
$$
H_0: X \sim P, \qquad \text{and} \qquad H_1: X \sim Q. 
$$
Let $\phi$ be a rejection rule that maps $X$ to $[0,1]$.  
Define the type I and II errors as 
$$
\alpha_{\phi} \coloneqq \mathbb{E}_P[\phi], \qquad \text{and} \qquad \beta_{\phi} \coloneqq \mathbb{E}_{Q} [1 - \phi]. 
$$

\noindent \cite{GDP} characterizes the hardness of a hypothesis testing problem between $P$ and $Q$ using its trade-off function, defined as follows. 

\begin{definition}[Trade-off Function, \cite{GDP}]
Let $P$ and $Q$ be two probability distributions on the same space. 
We define the  \textit{trade-off function} $T(P, Q):[0,1]\rightarrow[0,1]$ to be 
\[
T(P, Q)(x)=\inf\{\beta_\phi : \alpha_\phi \leqslant x\},
\]
where the infimum is taken over all measurable tests (rejection rules) $\phi$.
\end{definition}

\noindent Based on the notion of the trade-off function, \cite{GDP} formally defines $f$-differential privacy as follows:

\begin{definition}[$f$-Differential Privacy, \cite{GDP}] \label{definition-f-DP}
Let $f$ be a trade-off function. A randomized algorithm $\mathcal{A}$ is said to be $f$-differentially private ($f$-DP) if for every pair of neighboring datasets $(D,D')$, it satisfies
\[
T(\mathcal{A}(D), \mathcal{A}(D')) \geqslant f.
\]
Here, $\mathcal{A}(D)$ denotes the distribution induced by applying $A$ on the dataset $D$.  
\end{definition}

This definition indicates that an $f$-DP algorithm produces output distributions $\mathcal{A}(D)$ and $\mathcal{A}(D')$  that are difficult to distinguish by hypothesis testing, as specified by the trade-off function $f$. 

\subsection{Black-Box Auditing Point Differential Privacy}

Given a fixed algorithm $\mathcal{A}$ and a target trade-off function $f$, auditing whether $\mathcal{A}$ satisfies $f$-DP ideally requires verifying:
$$
T(\mathcal{A}(D), \mathcal{A}(D')) \geqslant f, \qquad \text{for all neighboring datasets }(D,D').
$$
While this verification is feasible if we have white-box access to the mechanism (i.e., access to internal workings of $\mathcal{A}$), real-world scenarios often limit auditors to black-box access—observing only the outputs of $\mathcal{A}$ for selected datasets.

Thus, in line with existing literature(~\cite{DPAuditing, Auditing3, Auditing-Impossibility}), we focus on auditing a simpler yet significant setting known as point differential privacy. 

\begin{definition}[Point $f$-Differential Privacy]
An algorithm $\mathcal{A}$ is \textit{point $f$-DP} for a specific pair of neighboring datasets $(D,D')$ if
\[
T(\mathcal{A}(D), \mathcal{A}(D')) \geqslant f.
\]
\end{definition}

Though more limited in scope, auditing point differential privacy serves as an essential building block toward auditing full differential privacy.

We formally frame the black-box auditing task as a statistical hypothesis testing problem. Fix an algorithm $\mathcal{A}$, a pair of neighboring datasets $(D,D')$, and a claimed trade-off function $f$. Given $n$ independent observations $(O_D^{(n)}, O_{D'}^{(n)})$ drawn from the output distributions $P = \mathcal{A}(D)$ and $P' = \mathcal{A}(D')$, the auditor seeks to test:

$$
H_0: T(P, P') \geqslant f, \qquad \text{versus} \qquad H_1: T(P, P') \ngeqslant f.  
$$

More formally, we have the following definition for an auditor. 
\begin{definition}[Auditing Mechanism]
An auditor $\pi = \pi(n,f,D,D')$ is a decision rule that maps $n$ samples $(O_D^{(n)}, O_{D'}^{(n)})\in \mathbb{R}^{n}\times\mathbb{R}^{n}$ to a binary indicator in $\{0,1\}$, where $0$ means that the auditor rejects the null hypothesis such $A$ obeys point $f$-DP.
\end{definition}

The mechanism $\pi$ effectively partitions the sample space $\mathbb{R}^{n}\times\mathbb{R}^{n}$ into:
\[
\pi^{DP}=\{(O_D^{(n)},O_{D'}^{(n)}): \pi(O_D^{(n)},O_{D'}^{(n)})=1\}, \quad \pi^{noDP}=(\pi^{DP})^c.
\]
Given that the auditor essentially tackles a hypothesis testing problem, we can define the usual type I and type II errors associated with the auditor. 

\begin{definition}[Type I and Type II Errors]
For an auditor $\pi=\pi(n,f,D,D')$, we define its type I and type II errors relative to a trade-off function $g$ as
\[
\mathrm{Error}_{I}(\pi,g)=\sup_{T(P,P')\geqslant g}\mathbb{P}\left((O_D^{(n)},O_{D'}^{(n)})\in \pi^{noDP}\right),
\]
\[
Error_{II}(\pi,g)=\sup_{T(P,P')\ngeqslant g}\mathbb{P}\left((O_D^{(n)},O_{D'}^{(n)})\in \pi^{DP}\right),
\]
where samples $(O_D^{(n)}, O_{D'}^{(n)})$ are drawn from $(P^{(n)},P'^{(n)})$.
\end{definition}

It is worth noting that the relative trade-off function $g$ is allowed to be different from the target trade-off function $f$ one aims to test. 
Clearly, we hope an auditor to have both small type I and type II errors relative to $g=f$. 

\section{A Conformal Inference-Based DP Auditing Mechanism}\label{sec:ConformalDP}

In this section, we propose a differential privacy (DP) auditing mechanism based on conformal inference. Our approach robustly controls the type I error rate even under fully black-box assumptions, requiring minimal restrictions on the underlying distributions.

The key insight behind Algorithm \ref{algorithm1} is to leverage conformal inference principles to derive high-probability upper and lower bounds for quantiles of the sampling distribution at each order statistic. These bounds facilitate distribution-free hypothesis tests, ensuring robust control of type I errors.

\begin{algorithm}[t]
\caption{CIPA: Conformal Inference point Privacy Auditing mechanism}\label{algorithm1}

\begin{enumerate}
    \item \textbf{Input:} Type I error control $\alpha\in (0,1)$, privacy function $f$, the number of output sampling $n$, the algorithm $\mathcal{A}$ and the neighboring dataset $(D,D')$.
    \item \textbf{Initialize:} $\varepsilon=\sqrt{\frac{-\ln\frac{\alpha}{4n}}{2n}}$, sampling the output dataset pair of $\mathcal{A}$: $(O_D^{(n)},O_{D'}^{(n)})$.
    \item \textbf{Sort} $(O_D^{(n)},O_{D'}^{(n)})$ in non-decreasing order:
    \[
    O_D^{(n)}=\{d_1,d_2,...,d_n\}, \quad O_{D'}^{(n)}=\{d'_1,d'_2,...,d'_n\}.
    \]
    \item \textbf{For} $k=1,2,...,n$, do:
    \begin{enumerate}
        \item \textbf{Compute}:
        \[
        l=card\left(\{l:d'_l<d_k\}\right), \quad l^{*}=n+1-card\left(\{l:d'_l>d_k\}\right).
        \]
        \item \textbf{If} 
        \[
        l>(n+1)\left(1-f\left(\frac{k}{n+1}+\varepsilon\right)+\varepsilon\right) \quad \text{or} \quad l^*<(n+1)\left(f\left(1-\frac{k}{n+1}+\varepsilon\right)-\varepsilon\right),
        \]
        \textbf{then:}
            \textbf{Output False and Exit this algorithm}.
    \end{enumerate}
    \item \textbf{Output True}.
\end{enumerate}

\end{algorithm}

To intuitively understand Algorithm \ref{algorithm1}, consider the following three steps:

\begin{enumerate}
    \item Step 1: Sort the sampled datasets such that $d_1\leqslant d_2\leqslant \dots\leqslant d_n$ and $d'_1\leqslant d'_2\leqslant \dots\leqslant d'_n$. For each index $1\leqslant k\leqslant n$, define hypothesis tests $\phi^1_k$ and $\phi^2_k$ with null hypothesis $H_0$: $x\sim P$, and rejection regions $R^1_k=(-\infty,d_k]$ and $R^2_k=[d_k,+\infty)$, respectively.

    \item Step 2: For test $\phi^1_k$, Theorem \ref{theorem-conformal} guarantees that, with high probability, the type I error satisfies $\mathbb{P}(O_D\in R^1_k)\geqslant \frac{k}{n+1} - \varepsilon$. Under $H_0$, the corresponding type II error satisfies
\[
\mathbb{P}(O_{D'} \in R^1_k) \geqslant f\left(\mathbb{P}(O_D\notin R^1_k)\right)\geqslant f\left(1-\frac{k}{n+1}+\varepsilon \right),
\]
which implies
\[
\mathbb{P}(O_{D'} \notin R^1_k)\leqslant 1-f\left(1-\frac{k}{n+1}+\varepsilon \right).
\]

\item Step 3: Using a similar reasoning, we derive a relationship between indices $l^*$ and $k$ with high probability. Specifically, combining
\[
\mathbb{P}(O_{D'} \notin R^1_k) \geqslant \mathbb{P}(O_{D'} \geqslant d'_{l^*}) \geqslant \frac{n+1-l^*}{n+1} - \varepsilon,
\]
with conformal inference results from Theorem \ref{theorem-conformal}, we ensure a robust relationship that holds distribution-free. Analogous analysis applies to $R^2_k$.

Iterating over all order statistics, we obtain a comprehensive auditing procedure ensuring robust type I error control.
\end{enumerate}

This procedure leads to the following theorem, establishing strong theoretical guarantees:

\begin{theorem}\label{theorem-typeI}
Let Algorithm \ref{algorithm1} run with any specified $\alpha \in(0,1)$. Without any further assumptions about the distributions $P$ and $P'$, Algorithm \ref{algorithm1} guarantees the type I error for point DP testing does not exceed $\alpha$, formally:
\[
\text{Error}_{I}\left(\pi(n,f,D,D'),f\right)\leqslant \alpha.
\]
\end{theorem}

\noindent See Appendix \ref{proof-theorem-typeI} for the proof. 
Our algorithm design is inspired by the classical result of training-conditional coverage in conformal prediction (Theorem \ref{theorem-conformal}, detailed in the appendix).

\subsection{Small Type II error is Impossible in the Distribution-Free Setting}
A natural question arises: does the proposed auditor $\pi$ have a small type II error? 
In other words, when the mechanism $\mathcal{A}$ is not differentially private, can the auditor correctly reject the null hypothesis based on the $n$ samples $(O_D^{(n)}, O_{D'}^{(n)})$?

It should be immediately clear that for any auditor $\pi$ with a small type I error, the worst-case type II error $Error_{II}(\pi,f)=\sup_{T(P,P')\ngeqslant f}\mathbb{P}\left((O_D^{(n)},O_{D'}^{(n)})\in \pi^{DP}\right)$ is large.  
The following example is instrumental to understand this simple lower bound. 

\begin{example}
    Consider two Gaussian distributions $P = \mathcal{N}(0,1)$, $Q = \mathcal{N}(1,1)$ with a trade-off function $T(P, Q) = f$.
    Let $Q' = \mathcal{N}(1+\zeta,1)$ with $\zeta>0$ be a perturbed version of $Q$.  
    Clearly, we have $T(P, Q') \ngeqslant f$. However,  $d_{TV}(Q, Q') \to 0$ as $\zeta$ approaches $0$.  
As a result, for any sample size $n$ and any auditor $\pi=\pi(n,f)$, we have:
$$
\mathbb{P}_{O_D^{(n)}\sim P^{(n)}, O_{D'}^{(n)}\sim Q'^{(n)}}\left((O_D^{(n)},O_{D'}^{(n)})\in \pi^{DP}\right)\geqslant \mathbb{P}_{O_D^{(n)}\sim P^{(n)}, O_{D'}^{(n)}\sim Q^{(n)}}\left((O_D^{(n)},O_{D'}^{(n)})\in \pi^{DP}\right)-o_\zeta(1),
$$
which further implies $Error_{II}(\pi(n,f), f)\geqslant 1-Error_{I}(\pi(n,f), f)$.
\end{example}

Therefore, in order to meaningfully discuss the power of an auditor, we need to assume that in the alternative hypothesis $H_1$, $T(P,P') \ngeqslant g$ for some relative trade-off function $g$ that is separated from $f$ that defines the null hypothesis. 
Such a choice of $g$ brings a necessary “buffer zone” between the null and alternative hypotheses. 

Nevertheless, in the following theorem, we provide a strong impossibility result: even when $g$ is drastically different from $f$, no auditor whatsoever with a small type I error can also have a small type II error.  

\begin{theorem}\label{theorem-Impossibility}
Fix a target trade-off function $f$. 
For any $n\geqslant1$ and any auditor $\pi=\pi(n,f)$, for any relative trade-off function $g:[0,1]\rightarrow[0,1]$ with $\|g\|_{\infty}=g(0) >0 $, we have 
$$
Error_{II}(\pi(n,f),g) \geqslant  1-Error_{I}(\pi(n,f),f). 
$$
\end{theorem}

\noindent See Appendix \ref{proof-theorem-Impossibility} for the proof. 

In particular, Theorem~\ref{theorem-Impossibility} states that if an auditing mechanism has a small Type I error, meaning it can correctly accept algorithms that satisfy the expected $f$-DP guarantee with high probability, then for any non-trivial trade-off function $g$ such that $||g||_{\infty}>0$, there exists an instance that does not satisfy $g$-DP but is erroneously accepted by the auditing mechanism with nearly the same high probability.
Hence, without further assumptions, no auditing mechanism can simultaneously have small type I and type II errors.

\subsection{Small Type II Error under MLR Assumption}

Now let us return to calculating the type II error of the proposed conformal inference-based auditor $\pi$. 
As we just discussed, if no further assumption is made, 
the type II error can not be made small. 
There necessary assumptions on the underlying distributions $P, P'$ are needed.

To this end, we introduce the following general assumption, based on which we can have a tight control of the type II of our auditor. 

\begin{assumption}[Monotone Likelihood Ratio] \label{MLR assumption}
We assume that the algorithm $\mathcal{A}$ operates within a class of random distributions that possess the monotone likelihood ratio (MLR) property, that is,
$$
\forall P,P'\in \mathcal{P}, \frac{f_P(x)}{f_{P'}(x)} \text{ is a monotone function for $x$ on $\{x\in \mathbb{R} \mid (f_{P}(x), f_{P'}(x))\neq(0,0)\}$}.
$$
Here we define $\frac{c}{0}=+\infty$ for $c>0$, and $f_P(x)$, $f_{P'}(x)$ represent the density function of the continuous distributions $P$ and $P'$ respectively.
\end{assumption}

\begin{remark}
    This assumption encompasses many common categories of differential privacy mechanisms. For instance, the widely used Gaussian distribution with fixed variance or Laplace distribution with fixed scale parameter falls under this assumption. Besides, Poisson, binomial, or generally, any regular exponential family with $g(t|\theta)=h(t)c(\theta)e^{w(\theta)t}$ has this property if $w(\theta)$ is a monotone function.
\end{remark}

With Assumption \ref{MLR assumption} in place, the hypothesis tests using one-sided intervals employed in Algorithm~\ref{algorithm1} are optimal. Consequently, this leads to tight control of the type II error.

\begin{theorem}\label{theorem-typeII}
Let Algorithm \ref{algorithm1} runs with any given $\alpha \in(0,1)$ and $n\geqslant 1$. For any trade-off function $g:[0,1]\rightarrow[0,1]$ satisfied $g<f$ in $[0,1)$, for any $r\in (0,1)$ denote $g_r=g\mathbbm{1}_{[0,r]}$ and
$$
\delta=\delta_r=\sup_{\delta\geqslant0}\{\forall 1\leqslant k\leqslant n:\  g_r(\max\{\frac{k-1}{n+1}-\delta,0\})\leqslant \max\{f(\min\{\frac{k}{n+1}+\varepsilon, 1\})-\frac{1}{n+1}-\varepsilon-\delta,0\}\}
$$
where $\varepsilon=\sqrt{\frac{-\ln\frac{\alpha}{4n}}{2n}}$ and define $\sup_{\delta\geqslant 0}\phi=0$ for the empty set $\phi$.
Then, under Assumption \ref{MLR assumption}, we have:
$$
Error_{II}(\pi(n,f,D,D'),g_r)\leqslant \min\{4ne^{-2n\delta^2},1\}
$$
where the upper bound on the right side approaches $0$ when $n\rightarrow +\infty$ for any fixed $r\in (0,1)$. Specifically, if $f$ or $g$ is Lipchitz continuous with Lipchitz parameter $L$, and denote $\gamma=\inf_{[0,r]} (f-g_r)>0$ is a positive constant, we have
$$
Error_{II}(\pi(n,f,D,D'),g_r)\leqslant O(4ne^{-\frac{n\gamma^2}{(L+1)^2}})=o(1)
$$
\end{theorem}
\noindent See Appendix \ref{proof-theorem-typeII} for the proof.

Here we introduce a truncation of the function $g$ near $1$, denoted as $g_r = g \mathbbm{1}_{[0,r]}$. However, this truncation is not a limitation imposed by our proof technique. In fact, the following example demonstrates that for any function $g$ that remains strictly positive on $[0,1)$, it is impossible to establish an effective control of the type II error.

\begin{example}
    Consider uniform distributions $P = U(0,1)$ with a trade-off function $T(P, P)(x) = 1-x\geqslant f(x)$ $\forall x\in [0,1]$.
    Let $Q = U(0,1-\zeta)$ with $\zeta\in (0,1)$ be a perturbed version of $P$.
    It is straightforward to check that $(P,Q)$ satisfies Assumption \ref{MLR assumption} and $T(P, Q)(x)=(1-\frac{x}{1-\zeta})\mathbbm{1}_{[0,1-\zeta)}(x)<g(x)$ in $[1-\zeta, 1)$ if $g>0$ in $[0,1)$. However,  $d_{TV}(P, Q) \to 0$ as $\zeta$ approaches $0$.  
As a result, for any sample size $n$ and any auditor $\pi=\pi(n,f)$, 
we have:
$$
\mathbb{P}_{O_D^{(n)}\sim P^{(n)}, O_{D'}^{(n)}\sim Q^{(n)}}\left((O_D^{(n)},O_{D'}^{(n)})\in \pi^{DP}\right)\geqslant \mathbb{P}_{O_D^{(n)}\sim P^{(n)}, O_{D'}^{(n)}\sim P^{(n)}}\left((O_D^{(n)},O_{D'}^{(n)})\in \pi^{DP}\right)-o_\zeta(1),
$$
which further implies $Error_{II}(\pi(n,f), g)\geqslant 1-Error_{I}(\pi(n,f), f)$.
\end{example}

\section{Constructing Confidence Bands for the Trade-Off Function}
In the end, we discuss a related problem to privacy auditing: constructing confidence bands for the trade-off function. 
Addressing this problem is useful when we do not have a specific privacy parameter, i.e., the trade-off function claimed by the algorithm developer. Instead, we aim to infer the privacy function based on the algorithm's outputs via constructing valid confidence bands for the algorithm's trade-off function.

To begin with, we have the following definition for the confidence band with level $\alpha$. 

\begin{definition}
(\textbf{Confidence Band for Point DP Function}).\\
Fix a mechanism $\mathcal{A}$, a pair of neighboring datasets $(D,D')$. 
Let $f \coloneqq T(\mathcal{A}(D), \mathcal{A}(D'))=T(P,P')$ be the unknown trade-off function. Given $n$ samples $O_D^{(n)}\sim P^{(n)}$ and $O_{D'}^{(n)}\sim P'^{(n)}$, two functions 
$f^{lower},  f^{upper}: [0,1] \to [0,1]$, measurable functions of the observations, is an $(1-\alpha)$-confidence band for the true trade-off function $f$ if for any admissible distribution pair $(P,P')$,
$$
\mathbb{P}_{O_D^{(n)}\sim P^{(n)}, O_{D'}^{(n)}\sim P'^{(n)}}\left(f\in[f^{lower}, f^{upper}]\right)\geqslant 1-\alpha, 
$$
where $f\in[f^{lower}, f^{upper}]$ means for $\forall x \in [0,1], f(x)\in[f^{lower}(x), f^{upper}(x)]$.
\end{definition}

\begin{algorithm}[t]
\caption{CIPB: Conformal Inference point Privacy confidence Bands constructor}\label{algorithm2}

\begin{enumerate}
    \item \textbf{Input:} Confidence level control $\alpha\in (0,1)$, the number of output sampling $n$, the algorithm $\mathcal{A}$ and the neighboring dataset $(D,D')$.
    \item \textbf{Initialize:} 
    $\varepsilon=\sqrt{\frac{-\ln\frac{\alpha}{8n}}{2n}}$,
    sampling the output dataset pair of $\mathcal{A}$: $(O_D^{(n)},O_{D'}^{(n)})$.
    \item \textbf{Sort} $(O_D^{(n)},O_{D'}^{(n)})$ in non-decreasing order:
    \[
    O_D^{(n)}=\{d_1,d_2,...,d_n\}, \quad O_{D'}^{(n)}=\{d'_1,d'_2,...,d'_n\}.
    \]
    \item \textbf{For} $k=1,2,...,n$, \textbf{compute:}
        \[
        l_k=card\left(\{l:d'_l<d_k\}\right), \quad l^*_k=n+1-card\left(\{l:d'_l>d_k\}\right).
        \]
    \item \textbf{For} $k=1,2,...,n$, \textbf{record:}
        \[
        \left(U_x(k), U_y(k)\right) =
        \left(\min\left\{\frac{k}{n+1}+\varepsilon, 1\right\}, \min\left\{\frac{n+1-l_k}{n+1}+\varepsilon, \frac{l^*_{n+1-k}}{n+1}+\varepsilon, 1\right\}\right).
        \]
        \[
        \left(L_x(k), L_y(k)\right) =
        \left(\max\left\{\frac{k}{n+1}-\varepsilon, 0\right\}, \max\left\{\min\{\frac{n+1-l^*_k}{n+1}-\varepsilon, \frac{l_{n+1-k}}{n+1}-\varepsilon\}, 0\right\}\right).
        \]
    \item \textbf{Output:} 
    \begin{align*}
    f^{upper}(x) &= \frac{U_y(k)-U_y(k+1)}{U_x(k)-U_x(k+1)}(x-U_x(k+1))+U_y(k+1), \quad \forall x\in [U_x(k), U_x(k+1)].\\
    f^{lower}(x) &= L_y(k+1), \quad \forall x\in (L_x(k), L_x(k+1)], \quad f^{lower}(0)=L_y(1).
    \end{align*}
    \textbf{where:}
    \[
    0\leqslant k\leqslant n, \quad U_x(0)=L_x(0)=0, \quad U_y(0)=1.
    \]
    \[
    (U_x(n+1), U_y(n+1))=(L_x(n+1), L_y(n+1))=(1, 0).
    \]
\end{enumerate}

\end{algorithm}

Following the idea as Algorithm \ref{algorithm1}, we propose the Algorithm \ref{algorithm2} based on conformal inference to construct the confidence band for point DP. And the algorithm can be roughly divided into the following three steps:

\begin{enumerate}
    \item Step 1: Sort the sampled data: $d_1\leqslant d_2\leqslant...\leqslant d_n$, $d'_1\leqslant d'_2\leqslant...\leqslant d'_n$. And for $1\leqslant k\leqslant n$, we perform the hypothesis test $\phi^1_k$ and $\phi^2_k$ with $H_0$: $x\sim P$ and rejection areas $R^1_k=(-\infty,d_k]$ and $R^2_k=[d_k,+\infty)$.
    \item Step 2: 
    For each test, following the same idea as in Step 2 of Algorithm \ref{algorithm1}, we can obtain the upper and lower bounds for type I and type II errors of each test. Consequently, we can derive the upper and lower bounds for a set of grid values, namely $f\left(U_x(k)\right) \leqslant U_y(k)$ and $f\left(L_x(k)\right) \leqslant L_y(k)$, with high probability. Here, the upper bounds are always valid, while the validity of the lower bounds depends on whether either $\phi^1_k$ or $\phi^2_k$ is an optimal test, which holds under Assumption \ref{MLR assumption}.

    \item Step 3: Finally, leveraging the monotonicity and convexity of the trade-off function (\cite{GDP}), we connect the grid points of the upper bounds into a piecewise linear curve and extend the grid points of the lower bounds to the entire region by taking the right endpoints. This yields high-probability upper and lower bands for the true $f$.
\end{enumerate}

We have the following performance guarantees for Algorithm~\ref{algorithm2}. 

\begin{theorem}\label{theorem-CI}
Let Algorithm \ref{algorithm2} runs with any given $\alpha \in(0,1)$. For any $n \geqslant 1$, $[0, f^{upper}]$ is an $(1-\frac{\alpha}{2})$-confidence band for the true trade-off function $f$.

If Assumption \ref{MLR assumption} holds, then $[f^{lower},f^{upper}]$ is an $1-\alpha$-confidence band for the true trade-off function $f$. 
\end{theorem}

Theorem \ref{theorem-CI} indicates that, under a fully black-box setting, we can effectively construct a confidence upper band for the true trade-off function. Moreover, when Assumption \ref{MLR assumption} holds, we can construct a valid confidence band for the trade-off function. 
In fact, similar to the impossibility result in Theorem \ref{theorem-Impossibility}, we can also establish the impossibility of providing valid confidence lower bands in the distribution-free setting. See Appendix \ref{appendix-impossibility-CI} for details.

More importantly, the confidence band we propose becomes increasingly tighter as the sample size $n$ increases, as seen from the following result. 

\begin{theorem} \label{theorem-asymtotic convergence}
    When Assumption~\ref{MLR assumption} holds, we have  
$$
\forall\delta>0, \lim_{n\rightarrow+\infty}\sup_{P,P'}\ \mathbb{P}_{O_D^{(n)}\sim P^{(n)}, O_{D'}^{(n)}\sim P'^{(n)}}\left(||\left(f^{upper}(x)-f^{lower}(x)\right)\Big{|}_{x\geqslant \sqrt{\frac{-\ln\frac{\alpha}{8n}}{2n}}+\frac{1}{n+1}}||_{\infty}>\delta\right)=0. 
$$
\end{theorem}

Combining the previous two theorems, one sees that under Assumption~\ref{MLR assumption}, both the upper and lower confidence bands converge to the true trade-off function, except on the point $f(0)$. 
This, in fact, is not a weakness of our construction. 
More specifically, one can prove that the true value of the trade-off function at $0$ $f(0)$ cannot be consistently estimated. This is because although the trade-off function $f$ is continuous with respect to the TV distance on $(0, 1]$, but it is discontinuous at $0$ with respect to the TV distance. Formally, we have the following example to illustrate it.

\begin{example}
    Consider two Gaussian distributions $P = \mathcal{N}(-1,\sigma^2)$ and $Q = \mathcal{N}(1,\sigma^2)$ with a trade-off function $T(P, Q) = f$. Let $P'=\mathcal{N}\left(-1,\sigma^2\right)\big{|}_{x\leqslant0}$ and $Q'=\mathcal{N}\left(1,\sigma^2\right)\big{|}_{x\geqslant0}$ are the conditional distributions of $P$ and $Q$ with a trade-off function $T(P', Q') = f'\equiv0$. 
    Clearly, $(P,Q,P',Q')$ satisfies Assumption \ref{MLR assumption} and $f(0)=1$, $f'(0)=0$. However, $d_{TV}(P, P')= d_{TV}(Q, Q')=\phi(-\frac{1}{\sigma^2})$ can be arbitrarily small when $\sigma^2$ goes to $0$. Here $\phi$ represents the CDF of the standard normal distribution $\mathcal{N}(0,1)$. As a result, for any sample size $n$ and any estimator $\pi=\pi(n)$, the TV distance between the estimations of $f$ and $f'$, $d_{TV}\left(\hat{f}^\pi(0), \hat{f}'^\pi(0)\right)$, can be arbitrarily small.
\end{example}

This conclusion means that, for any $n$, there exist two sets of distributions with radically different values at $f(x)|_{x=0}$, but the data distributions they generate are almost identical. Therefore, for any large $n$, it is impossible to accurately estimate $f(0)$.

\section{Conclusions and discussions}
This papers studies the privacy auditing problem in the black-box setting. 
We formulate this problem using the language of statistical hypothesis testing. 
We propose a conformal-inference-based auditing method CIPA that has a small type I error in the finite-sample setting. 
Turning to the type II error, we show that without further assumptions, simultaneously small type I and II errors are not achievable for any auditing method. 
Nevertheless, under a general monotone likelihood ratio assumption, we prove that our auditing mechanism can simultaneously control the type I and type II errors.
We also extend our method to construct valid confidence bands for the trade-off function in the finite-sample case. 

Our study leaves open quite a few interesting directions to pursue. First, are there any other reasonable assumptions one can make to allow a tight control of both type I and II errors? Second, the main focus of this paper is to audit point differential privacy where the neighboring datasets are given and fixed. How to formulate the problem of auditing the true differential privacy is an important open question. 

\bibliographystyle{chicago}
\bibliography{citation}

\appendix

\section{Proof of Theorem \ref{theorem-typeI}} \label{proof-theorem-typeI}
We first state the classical training-conditional conformal prediction result below, which will be used in our proofs frequently.
\begin{theorem} (Theorem 4.1, \cite{Conformal-Notes}) \label{theorem-conformal}
Suppose the data points $X$ and $\{X_i\}_{i=1}^{n}$ are i.i.d., denote by $\{X_{(i)}\}_{i=1}^{n}$ the order statistics of $\{X_i\}_{i=1}^{n}$. Then for $1\leqslant k\leqslant n$ and for $\mathcal{C}=(-\infty, X_{(k)}]$ or $\mathcal{C}=[X_{(n-k+1)},+\infty)$ the conditional coverage $\mathbb{P}\left(X \in \mathcal{C}\mid \{X_i\}_{i=1}^{n}\right)$ stochastically dominates the $\operatorname{Beta}(k, n+1-k)$ distribution, and in particular, for any $\Delta>0$,
$$
\mathbb{P}\left(\mathbb{P}\left(X \in \mathcal{C} \mid \{X_i\}_{i=1}^{n}\right) \leq \frac{k}{n+1}-\Delta\right) \leq F_{\operatorname{Beta}(k, n+1-k)}(\frac{k}{n+1}-\Delta) \leq e^{-2 n \Delta^2}
$$
where $F_{\operatorname{Beta}(a, b)}$ denotes the CDF of the $\operatorname{Beta}(a, b)$ distribution.
\end{theorem}

For any $1\leqslant k\leqslant n$, by Theorem \ref{theorem-conformal}, for $\mathcal{C}=(-\infty,d_k]$ or $\mathcal{C}=[d_{n-k+1}, +\infty)$, we have
$$
\mathbb{P}\left(\mathbb{P}\left(O_D \in \mathcal{C} \mid O^{(n)}_D\right) \leqslant \frac{k}{n+1}-\varepsilon\right) \leqslant e^{-2 n \varepsilon^2}=\frac{\alpha}{4n},
$$
where $O_D$ is an independent draw from $P$. 
Thus, for a given $k$ and $\mathcal{C}=(-\infty,d_k]$ or $\mathcal{C}=[d_{n-k+1}, +\infty)$, with probability at least $ 1-\frac{\alpha}{4n}$, $\mathbb{P}\left(O_D \notin \mathcal{C} \right) \leqslant 1-\frac{k}{n+1}+\varepsilon$.

Under the null, i.e., $\mathcal{A}$ is point $f$-DP on $D$ and $D'$, we know that on the same high probability event as above, we have $\mathbb{P}\left(O_{D'} \in \mathcal{C} \right) \geqslant f(1-\frac{k}{n+1}+\varepsilon)$,  where $O_{D'}$ is an independent draw from $P'$.

Specifically, for $\mathcal{C}=(-\infty,d_k]$, we have with probability $\geqslant 1-\frac{\alpha}{4n}$, $\mathbb{P}\left(O_{D'} \leqslant d_k \right) \geqslant f(1-\frac{k}{n+1}+\varepsilon)$

Similarly, by Theorem \ref{theorem-conformal}, with probability at least $ 1-\frac{\alpha}{4n}$, $\mathbb{P}\left(O_{D'} \geqslant d'_{l^*} \right) \geqslant \frac{n+1-l^*}{n+1}-\varepsilon$. Recall that  $d'_{n+1}=+\infty$.

Note that by definition, we have  $d'_{l^*}>d_k$. 
Consequently, we have with probability at least $ 1-\frac{\alpha}{2n}$, $1\geqslant \mathbb{P}\left(O_{D'} \leqslant d_k \right)+\mathbb{P}\left(O_{D'} \geqslant d'_{l+1} \right)\geqslant f(1-\frac{k}{n+1}+\varepsilon)+\frac{n+1-l^*}{n+1}-\varepsilon$, which implies $l^*\geqslant (n+1)\left(f(1-\frac{k}{n+1}+\varepsilon)-\varepsilon\right)$.

Similarly, consider $\mathcal{C}=[d_k, +\infty)$, we have with probability $\geqslant 1-\frac{\alpha}{4n}$, $\mathbb{P}\left(O_{D'} \geqslant d_k \right) \geqslant f(\frac{k}{n+1}+\varepsilon)$. On the other hand, with probability $\geqslant 1-\frac{\alpha}{4n}$, $\mathbb{P}\left(O_{D'} \leqslant d'_l \right) \geqslant \frac{l}{n+1}-\varepsilon$ (define $d'_0=-\infty$). Thus we have $1\geqslant \mathbb{P}\left(O_{D'} \geqslant d_k \right)+\mathbb{P}\left(O_{D'} \leqslant d'_l \right)\geqslant f(\frac{k}{n+1}+\varepsilon)+\frac{l}{n+1}-\varepsilon$, which implies $l\leqslant(n+1)\left(1-f(\frac{k}{n+1}+\varepsilon)+\varepsilon\right)$.

In all, for any given $1\leqslant k\leqslant n$, under the null, with probability $\geqslant 1-\frac{\alpha}{n}$, we have
$l\leqslant(n+1)\left(1-f(\frac{k}{n+1}+\varepsilon)+\varepsilon\right)$ and $l^*\geqslant(n+1)\left(f(1-\frac{k}{n+1}+\varepsilon)-\varepsilon\right)$. 
This together with the union bound 
shows that Algorithm \ref{algorithm1} has a type I error at most $\alpha$.

\section{Proof of Theorem \ref{theorem-Impossibility}} \label{proof-theorem-Impossibility}
Let $Q=\text{Uniform}([0,1))$.  
Let $m$ be an integer that is much larger than $n$. For each $1\leq i \leq m^2$, let  $Q_i=\text{Uniform}(I_i)$ where $I_i=[\frac{i-1}{m^2}, \frac{i}{m^2})$.  
Moreover, define $\{P_j\}_{1\leqslant j\leqslant\tbinom{m^2}{m}}$ the set of distributions with the form $\frac{1}{m}\sum_{1\leqslant k\leqslant m} Q_{i_k}$, where $\{i_k\}_{1\leq k\leq m}$ are disjoint indices in $[m^2]$.  
In words, each $P_j$ is a uniform distribution on the union of selected $m$ intervals among all $m^2$ intervals $I_i$. Furthermore, we define the mixture distribution  $Q_*^{(n)}=\tbinom{m^2}{m}^{-1}\sum P_j^{(n)}$, where $P_j^{(n)}$ represent $P_j$ multiply themselves by $n$ times.

As shown in the lemma below, the two distributions $Q$ and $Q_*$ are quite similar. 

\begin{lemma}\label{lem:small_TV}
$d_{TV}(Q^{(n)},Q_*^{(n)})=o(1).$ 
\end{lemma}

Thus, by definition, $d_{TV}(Q^{(n)},Q_*^{(n)})=o(1)$ can be arbitrarily small for large $m$, that is, for any $\zeta>0$ we can choose $m$ large enough to satisfy:
$$
\mathbb{P}_{O_D^{(n)}\sim Q^{(n)}, O_{D'}^{(n)}\sim Q_*^{(n)}}((O_D^{(n)},O_{D'}^{(n)})\in \pi^{DP})\geqslant \mathbb{P}_{O_D^{(n)}\sim Q^{(n)}, O_{D'}^{(n)}\sim Q^{(n)}}((O_D^{(n)},O_{D'}^{(n)})\in \pi^{DP})-\zeta.
$$
Note that $Q_*^{(n)}=\tbinom{m^2}{m}^{-1}\sum P^{(n)}_j$. Hence
$$
\mathbb{P}_{O_D^{(n)}\sim Q^{(n)}, O_{D'}^{(n)}\sim Q_*^{(n)}}((O_D^{(n)},O_{D'}^{(n)})\in \pi^{DP})=\tbinom{m^2}{m}^{-1}\sum \mathbb{P}_{O_D^{(n)}\sim Q^{(n)}, O_{D'}^{(n)}\sim P^{(n)}_j}((O_D^{(n)},O_{D'}^{(n)})\in \pi^{DP}),
$$
which implies the existence of $P\in \{P_j\}$ such that
$$
\mathbb{P}_{O_D^{(n)}\sim Q^{(n)}, O_{D'}^{(n)}\sim P^{(n)}}((O_D^{(n)},O_{D'}^{(n)})\in \pi^{DP})\geqslant \mathbb{P}_{O_D^{(n)}\sim Q^{(n)}, O_{D'}^{(n)}\sim Q^{(n)}}((O_D^{(n)},O_{D'}^{(n)})\in \pi^{DP})-\zeta.
$$

And we simultaneously have $||T(P,Q)||_{\infty}=T(P,Q)(0)=\frac{m}{m^2}=\frac{1}{m}<g(0)$ for large $m$. Therefore, $T(P,Q)\ngeqslant g$, so we have $Error_{II}(\pi(n,f),g) \geqslant  1-Error_{I}(\pi(n,f),f)-\zeta$ for any $\zeta>0$, i.e. $Error_{II}(\pi(n,f),g) \geqslant  1-Error_{I}(\pi(n,f),f)$.

We finished the proof.

\begin{proof}[Proof of Lemma~\ref{lem:small_TV}]
Also, for $x\in[0,1)$, define a mapping function $I(x)=i$, where $x\in I_i$.

First, we want to prove that $d_{TV}(Q^{(n)}, Q_*^{(n)})$ can be arbitrarily small when $m$ is large enough. In fact, for a sampled data $x^{(n)}=(x_1,x_2,...,x_n)\in[0,1)^n$, we consider that the probability densities of $Q^{(n)}$ and $Q_*^{(n)}$ for this data condition on $\{I(x_i)\}$ is different from each other.

For $\{I(x_i)\}$ is not different from each other, the conditional probability is 0 for both $Q^{(n)}$ and $Q_*^{(n)}$. For $\{I(x_i)\}$ is different from each other, the conditional density in $Q^{(n)}(x^{(n)})$ is $q=1\times\frac{m^2}{m^2-1}\times...\times\frac{m^2}{m^2-n+1}$, and the conditional density in $Q_*^{(n)}(x^{(n)})$ is $q_*=\tbinom{m^2}{m}^{-1}\tbinom{m^2-n}{m-n}m^n\left(1\times \frac{m}{m-1}\times...\times \frac{m}{m-n+1}\right)=\frac{m^{2n}}{m^2(m^2-1)...(m^2-n+1)}=q$.

This means, $Q^{(n)}$ is exactly the same with $Q_*^{(n)}$ condition on $\{I(x_i)\}$ is different from each other. Denote event $E$ is $\{I(x_i)\}$ is different from each other, then it's easy to check that $Q^{(n)}(E)>1-\frac{1}{m^2}\tbinom{n}{2}>Q_*^{(n)}(E)\approx 1-\frac{1}{m}\tbinom{n}{2}$ for large $m$.

Therefore, for any event $A$,
\begin{align*}
    Q^{(n)}(A)-Q_*^{(n)}(A)&=Q^{(n)}(A\cap E|E)Q^{(n)}(E)-Q_*^{(n)}(A\cap E|E)Q_*^{(n)}(E)+Q^{(n)}(A\cap E^c)-Q_*^{(n)}(A\cap E^c)\\
    &\leqslant Q^{(n)}(A\cap E|E)(Q^{(n)}(E)-Q_*^{(n)}(E))+Q^{(n)}(E^c)\\
    &\leqslant Q^{(n)}(E)-Q_*^{(n)}(E)+Q^{(n)}(E^c)\\
    &=1-Q_*^{(n)}(E)\\
    &=O(\frac{1}{m}\tbinom{n}{2})\\
    &=o(1).
\end{align*}
\end{proof}

\section{Proof of Theorem \ref{theorem-typeII}} \label{proof-theorem-typeII}
We follow the idea used in the proof of Theorem \ref{theorem-typeI}. Below, let $T$ be the true trade-off function between $P$ and $P'$. Denote $\beta=4ne^{-2n\delta^2}$. The specified type II error is controlled by $1$ is obvious, we only need to show that the type II error is less than or equal to $\beta$. Suppose $\beta<1$, which means $\delta>0$.

Then $\delta=\sqrt{\frac{-\ln \frac{\beta}{4n}}{2n}}$. We define $g_n$ as follows:
$$
a_k=\frac{k}{n+1}-\delta\quad (\forall 1\leqslant k\leqslant n)\quad a_{n+1}=1 \quad a_0=-\infty
$$
$$
\forall x\in [0,1]\ \text{ that }  a_{k-1}<x\leqslant a_k,\  g_n(x)=\max\{f\left(\min\{\frac{k}{n+1}+\varepsilon,\ 1\}\right)-\varepsilon-\delta-\frac{1}{n+1},\ 0\}
$$

Roughly speaking, we define the grid value for $g_n$ that $g_n\left(\frac{k}{n+1}-\delta\right)=f\left(\frac{k}{n+1}+\varepsilon\right)-\varepsilon-\delta-\frac{1}{n+1}$ and set all $g_n(x)=g_n(\frac{k}{n+1}-\delta)$ for the minimal $k$ that $x\leqslant \frac{k}{n+1}-\delta$. Then, by definition, $g_r\leqslant g_n$ in $[0,1]$. We only need to prove that $Error_{II}(\pi, g_n)\leqslant \beta$.

Notice that $P$ and $P'$ are continuous distributions (Assumption \ref{MLR assumption}), with probability 1, $l^*=l+1$ for all $k$ in Algorithm \ref{algorithm1}.

We consider the case that $T\ngeqslant g_n$ and the algorithm output \textbf{True} to evaluate the type II error. By the definition of $g_n$ and $T$ is non-increasing, $T\ngeqslant g_n$ means that $\exists 1\leqslant k\leqslant n$, $T(a_k)< g(a_k)$ for $a_k=\frac{k}{n+1}-\delta$.

Same as the previous proof of the Theorem \ref{theorem-typeI}, by using the Theorem \ref{theorem-conformal}, with probability at least $1-\beta$, for all possible $\mathcal{C}=(-\infty,d_k]$ or $\mathcal{C}=[d_{n-k+1}, +\infty)$, $\mathbb{P}\left(O_D \in \mathcal{C} \right) \geqslant \frac{k}{n+1}-\delta$. Besides, for all possible corresponding $\mathcal{C'}=(-\infty,d'_l]$ (and the $n$ intervals on the other side), $\mathbb{P}\left(O'_D \notin \mathcal{C} \right) \geqslant 1-\frac{l}{n+1}-\delta$.

Because we assume $P$ and $P'$ has the MLR property \ref{MLR assumption}, by Neyman-Pearson lemma, neither $\mathcal{C}=(-\infty,d_k]$ or $\mathcal{C}=[d_{n-k+1}, +\infty)$ satisfies $(\mathcal{C},\ \mathbb{R}\backslash \mathcal{C})$ for $(P,P')$ are the optimal test strategy, this means for this particular $\mathcal{C}$, $\mathbb{P}\left(O_{D'} \notin \mathcal{C} \right)=T\left(\mathbb{P}\left(O_{D} \in \mathcal{C} \right)\right)$.

Thus with probability at least $1-\beta$, for all $1\leqslant k\leqslant n$, and for one of $\mathcal{C}=(-\infty,d_k]$ and $\mathcal{C}=[d_{n-k+1}, +\infty)$, we have:
$$
T(a_k)\geqslant T\left(\mathbb{P}\left(O_{D} \in \mathcal{C} \right)\right)=\mathbb{P}\left(O_{D'} \notin \mathcal{C} \right)
$$

Specifically, for the $k$ we previously mentioned, i.e. $T(a_k)< g(a_k)$, if $\mathcal{C}=(-\infty,d_k]$ satisfy the above inequation, since the algorithm output \textbf{True}, we know $l\leqslant (n+1)\left(1-f(\frac{k}{n+1}+\varepsilon)+\varepsilon\right)$, thus:
$$
g_n(a_k)> T(a_k)\geqslant T\left(\mathbb{P}\left(O_{D} \in \mathcal{C} \right)\right)=\mathbb{P}\left(O_{D'} \notin \mathcal{C} \right)\geqslant \frac{n-l}{n+1}-\delta\geqslant f\left(\frac{k}{n+1}+\varepsilon\right)-\varepsilon-\delta-\frac{1}{n+1}=g_n(a_k)
$$
This is a contradiction.

Similarly, for $\mathcal{C}=[d_{n-k+1}, +\infty)$, we denote $l'=argmax\{l:d'_l\leqslant d_{n-k+1}\}$, then $l'\geqslant (n+1)\left(f(\frac{k}{n+1}+\varepsilon)-\varepsilon\right)-1$:
$$
g_n(a_k)>T(a_k)\geqslant\frac{l'}{n+1}-\delta\geqslant f\left(\frac{k}{n+1}+\varepsilon\right)-\varepsilon-\delta-\frac{1}{n+1}=g_n(a_k)
$$

Therefore, we have the type II error $Error_{II}(\pi(n,f,D,D'),g_n)\leqslant\beta$.

Finally, for any fixed $r<1$, when $n$ goes to infinity, we want to show that $\beta$ goes to $0$.

In fact, $g<f$ in $[0,1)$ and $g$, $f$ are both continuous (by proposition 2 in \cite{GDP}), we know that $\gamma=\inf_{[0,r]} (f-g_r)>0$. Since $f$ is continuous uniformly in $[0,1]$, we have 
$$
\exists N, \forall |x_1-x_2|\leqslant\frac{1}{N+1}+\varepsilon+N^{-\frac{1}{3}}, |f(x_1)-f(x_2)|<\frac{\gamma}{2}<\gamma-\frac{1}{N+1}-N^{-\frac{1}{3}}
$$

Then, by definition, $\forall n\geqslant N$, $\delta=N^{-\frac{1}{3}}$ satisfy the inequality in Theorem \ref{theorem-typeII}, thus $\delta\geqslant N^{-\frac{1}{3}}$. Therefore, $\beta=4ne^{-2n\delta^2}\leqslant4ne^{-2nN^{-\frac{2}{3}}}$ goes to $0$.

If $f$ or $g$ is Lipchitz continuous, for sufficiently large $n$, we can check that $\delta\geqslant \frac{\gamma}{\sqrt{2}(L+1)}$ as following:

For $\delta_0=\frac{\gamma}{\sqrt{2}(L+1)}$ and $\frac{k-1}{n+1}-\delta_0\leqslant r$, $\frac{k}{n+1}+\varepsilon\leqslant r+o(\frac{1}{\sqrt{n}})$, thus $\gamma_0=\inf_{[0,r+o(\frac{1}{\sqrt{n}})]}(f-g)\geqslant 0.99\gamma$ for sufficiently large $n$. Therefore, 
\begin{align*}
    f(\min\{\frac{k}{n+1}+\varepsilon,1\})-g_r(\max\{\frac{k-1}{n+1}-\delta_0,0\})&\geqslant
    \gamma_0-L(\delta_0+\varepsilon+\frac{1}{n+1})\\
    &\geqslant (0.99-\frac{1}{\sqrt{2}})\gamma+\delta_0-o(\frac{L}{\sqrt{n}})\\
    &\geqslant \frac{1}{n+1}+\varepsilon+\delta_0
\end{align*}

the last inequality holds for sufficiently large $n$.

Therefore, with $\delta\geqslant \frac{\gamma}{\sqrt{2}(L+1)}$, we can obtain the result in the theorem.

\section{Proof of Theorem \ref{theorem-CI} and Theorem \ref{theorem-asymtotic convergence}} \label{proof-theorem-CI}
First, by the same procedure in the proof of Theorem \ref{theorem-typeI}, with probability $\geqslant 1-\frac{\alpha}{2}$, for any $1\leqslant k\leqslant n$, $f(U_x(k))\leqslant U_y(k)$, and of course, $f(0)\leqslant1$. By proposition 2.2 in \cite{GDP}, $f$ is convex, continuous and non-increasing, we know $f\leqslant f^{upper}$, where $f^{upper}$ is the polyline that connects these points.

Similarly, by the same procedure in the proof Theorem \ref{theorem-typeII}, under Assumption \ref{MLR assumption}, with probability $\geqslant 1-\frac{\alpha}{2}$, for any $1\leqslant k\leqslant n$, $f(L_x(k))\geqslant L_y(k)$ and of course, $f(1)=0$. Thus we have $f\geqslant f^{lower}$. Combine these two, with probability $\geqslant 1-\alpha$, $f\in [f^{lower}, f^{upper}]$.

Next we will show the asymptotic convergence in probability.

Under Assumption \ref{MLR assumption}, with probability 1, $l^*_k=l_k+1$ for all $k$.

For the given form of $f^{upper}$ and $f^{lower}$, we know that
$$
\forall x\geqslant\frac{1}{n+1}+\varepsilon=o(\frac{1}{\sqrt{n}}),\ \text{suppose } \frac{k}{n+1}+\varepsilon\leqslant x\leqslant \frac{k+1}{n+1}+\varepsilon, \text{ and } \frac{k'}{n+1}-\varepsilon< x\leqslant \frac{k'+1}{n+1}-\varepsilon
$$
$\text{ then } k'-k=O(2n\varepsilon)=\widetilde{O}(\sqrt{n}).$

Thus, by the definition of $f^{upper}$ and $f^{lower}$, we have
\begin{align*}
    |f^{upper}(x)-f^{lower}(x)|&\leqslant U_y(k)-L_y(k'+1)\\
    &\leqslant \max\{\frac{l_{k'+1}-l_k}{n+1}, \frac{l_{n+1-k}-l_{n-k'}}{n+1}\}+\widetilde{O}(\frac{1}{\sqrt{n}})\\
    &\leqslant \frac{\max_{k'-k=\widetilde{O}(\sqrt{n})}\{l_{k'}-l_k\}}{n+1}+\widetilde{O}(\frac{1}{\sqrt{n}})
\end{align*}

According to the Assumption \ref{MLR assumption}, without losing any generality, we assume $dP/dP'$ decreases. By using the Theorem \ref{theorem-conformal}, with probability $\geqslant 1-4ne^{-2\delta^2n^{\frac{1}{3}}}$, for any $1\leqslant k\leqslant n$, $|\frac{l_k}{n+1}-\mathbb{P}(O'_D\leqslant d_k)|\leqslant \delta n^{-\frac{1}{3}}$, and $\mathbb{P}(O'_D\leqslant d_k)=f(\mathbb{P}(O_D> d_k))\in \left[f\left(1-\frac{k}{n+1}+\delta n^{-\frac{1}{3}}\right),\ f\left(1-\frac{k}{n+1}-\delta n^{-\frac{1}{3}}\right)\right]$.

Therefore, with probability $\geqslant 1-4ne^{-2\delta^2n^{\frac{1}{3}}}$,
$$
\frac{\max_{k'-k=\widetilde{O}(\sqrt{n})}\{l_{k'}-l_k\}}{n+1}\leqslant 2\delta n^{-\frac{1}{3}}+\sup_{|x-x'|\leqslant 2\delta n^{-\frac{1}{3}}+\widetilde{O}(\frac{1}{\sqrt{n}})} |f(x')-f(x)|=o(1)
$$
since $f$ is (uniformly) continuous in $[0,1]$.

Thus for a fixed $\delta>0$, with the large $n$ satisfied that $2\delta n^{-\frac{1}{3}}+\sup_{|x-x'|\leqslant 2\delta n^{-\frac{1}{3}}+\widetilde{O}(\frac{1}{\sqrt{n}})} |f(x')-f(x)|<\delta$, for any $P,P'\in \mathcal{P}$, we have:

$$
\mathbb{P}_{O_D^{(n)}\sim P^{(n)}, O_{D'}^{(n)}\sim P'^{(n)}}\left(||\left(f^{upper}-f^{lower}\right)\Big{|}_{x\geqslant \sqrt{\frac{-\ln\frac{\alpha}{8n}}{2n}}+\frac{1}{n+1}}||_{\infty}>\delta\right)\leqslant 4ne^{-2\delta^2n^{\frac{1}{3}}}=o(1)
$$

Thus we finished the proof.

\section{Impossibility of Providing Valid Confidence Lower Bands in the Distribution-free Setting} \label{appendix-impossibility-CI}

\begin{theorem}\label{theorem-Impossibility-CI}
For any $n\geqslant1$ and any confidence lower band constructor $\pi=\pi\left(n,D,D'\right)$ and $\varepsilon,\delta>0$, there exists two continuous probability distributions $P$ and $Q$, that satisfied $||T(P,Q)||_{\infty}\leqslant \varepsilon$ and:
$$
\mathbb{P}_{O_D^{(n)}\sim P^{(n)}, O_{D'}^{(n)}\sim Q^{(n)}}\left(||f^{lower}||_{\infty}>\varepsilon\right)\geqslant \mathbb{P}_{O_D^{(n)}\sim P^{(n)}, O_{D'}^{(n)}\sim P^{(n)}}\left(||f^{lower}||_{\infty}>\varepsilon\right)-\delta
$$
\end{theorem}

Likewise, this theorem indicates that under the fully black-box assumption, it is impossible to establish any meaningful lower bands. Concretely, if a mechanism can establish a valid lower band with high probability, then the error probability represented by the left-hand side of the inequality in theorem \ref{theorem-Impossibility-CI} should be very small. However, this implies that the lower band established by the mechanism for two identical distributions (whose trade-off function is $f(x) = 1 - x$) would with high probability satisfy $||f^{lower}(x)||_{\infty} \leqslant \varepsilon$. Such a lower band is almost meaningless. In other words, under this circumstance, it is impossible to establish any meaningful lower band with high probability.

We provide the following proof.

\begin{proof}
We define $Q$, $Q_*^{(n)}$ as in the previous proof of Theorem \ref{theorem-Impossibility} in Appendix \ref{proof-theorem-Impossibility}, so that $d_{TV}(Q^{(n)},Q_*^{(n)})=o(1)$ can be arbitrarily small for large $m$, that is, we can choose $m$ large enough to satisfy:
$$
\mathbb{P}_{O_D^{(n)}\sim Q^{(n)}, O_{D'}^{(n)}\sim Q_*^{(n)}}(||f^{lower}||_{\infty}>\varepsilon)\geqslant \mathbb{P}_{O_D^{(n)}\sim Q^{(n)}, O_{D'}^{(n)}\sim Q^{(n)}}(||f^{lower}||_{\infty}>\varepsilon)-\delta.
$$

Noticed that $Q_*^{(n)}=\tbinom{m^2}{m}^{-1}\sum P^{(n)}_j$, hence
$$
\mathbb{P}_{O_D^{(n)}\sim Q^{(n)}, O_{D'}^{(n)}\sim Q_*^{(n)}}(||f^{lower}||_{\infty}>\varepsilon)=\tbinom{m^2}{m}^{-1}\sum \mathbb{P}_{O_D^{(n)}\sim Q^{(n)}, O_{D'}^{(n)}\sim P^{(n)}_j}(||f^{lower}||_{\infty}>\varepsilon).
$$

Therefore, $\exists P\in \{P_j\}$, such that
$$
\mathbb{P}_{O_D^{(n)}\sim Q^{(n)}, O_{D'}^{(n)}\sim P^{(n)}}(||f^{lower}||_{\infty}>\varepsilon)\geqslant \mathbb{P}_{O_D^{(n)}\sim Q^{(n)}, O_{D'}^{(n)}\sim Q^{(n)}}(||f^{lower}||_{\infty}>\varepsilon)-\delta.
$$

And we simultaneously have $||T(P,Q)||_{\infty}=T(P,Q)(0)=\frac{m}{m^2}=\frac{1}{m}\leqslant\varepsilon$ for large $m$.

We finished the proof.
\end{proof}

\end{document}